\documentclass[acmlarge]{acmart}
\makeatletter
\newcommand{\myconfshort}{\acmConference@shortname}
\newcommand{\myconffull}{\acmConference@name}
\newcommand{\myconfdate}{\acmConference@date}
\newcommand{\myconfloc}{\acmConference@venue}
\AtBeginDocument{
  \fancypagestyle{firstpagestyle}{
    \fancyhead{}%
    \fancyfoot[C]{}%
  }
  \fancyhf{}
  \fancyhead[LO]{\@headfootfont\shorttitle}%
  \fancyhead[RE]{\@headfootfont\@shortauthors}%
  \fancyhead[LE]{\@headfootfont\footnotesize \myconfshort, \myconfdate, \myconfloc}%
  \fancyhead[RO]{\@headfootfont\footnotesize \myconfshort, \myconfdate, \myconfloc}%
  \fancyfoot[C]{}%
}
\makeatother
\acmBooktitle{\conffull\@ (\confshort), \confdate, \confloc}
\AtBeginDocument{%
  }

\setcopyright{acmlicensed}
\copyrightyear{2026}
\acmYear{2026}
\setcopyright{cc}
\setcctype{by}
\acmConference[FAccT '26]{The 2026 ACM Conference on Fairness, Accountability, and Transparency}{June 25--28, 2026}{Montreal, QC, Canada}
\acmBooktitle{The 2026 ACM Conference on Fairness, Accountability, and Transparency (FAccT '26), June 25--28, 2026, Montreal, QC, Canada}
\acmDOI{10.1145/3805689.3806759}
\acmISBN{979-8-4007-2596-8/2026/06}

\usepackage{color-edits}
\addauthor{bk}{blue}
\addauthor{hh}{red}
\addauthor{rk}{cyan}
\addauthor{kh}{violet}
\addauthor{rg}{magenta}

\newcommand{\xhdr}[1]{\vspace{1.7mm}\noindent{{\bf #1.}}}

\usepackage{pdfpages}
\usepackage{subcaption}

\begin{document}

\title[Toward Third-Party Assurance of AI Systems]{Toward Third-Party Assurance of AI Systems: Design Requirements, Prototype, and Early Testing}

\author{Rachel M. Kim}
\authornote{Co-first authors contributed equally to this research.}
\email{rachelmkim@cmu.edu}
\orcid{0009-0000-3325-243X}
\affiliation{%
  \institution{Carnegie Mellon University}
  \city{Pittsburgh}
  \state{Pennsylvania}
  \country{USA}
}
\author{Blaine Kuehnert}
\authornotemark[1]
\email{blainekuehnert@cmu.edu}
\orcid{0009-0001-5702-5693}
\affiliation{%
  \institution{Carnegie Mellon University}
  \city{Pittsburgh}
  \state{Pennsylvania}
  \country{USA}
}
\author{Alice Lai}
\email{alicelai@cmu.edu}
\orcid{0009-0009-0352-2596}
\affiliation{%
  \institution{Carnegie Mellon University}
  \city{Pittsburgh}
  \state{Pennsylvania}
  \country{USA}
}
\author{Kenneth Holstein}
\email{kjholste@cs.cmu.edu}
\orcid{0000-0001-6730-922X}
\affiliation{%
  \institution{Carnegie Mellon University}
  \city{Pittsburgh}
  \state{Pennsylvania}
  \country{USA}
}
\author{Hoda Heidari}
\authornote{Co-last authors contributed equally to this research.}
\email{hheidari@cmu.edu}
\orcid{0000-0003-3710-4076}
\affiliation{%
  \institution{Carnegie Mellon University}
  \city{Pittsburgh}
  \state{Pennsylvania}
  \country{USA}
}
\author{Rayid Ghani}
\authornotemark[2]
\email{rayid@cmu.edu}
\orcid{0000-0003-0235-1843}
\affiliation{%
  \institution{Carnegie Mellon University}
  \city{Pittsburgh}
  \state{Pennsylvania}
  \country{USA}
}

\renewcommand{\shortauthors}{Rachel M. Kim et al.}

\begin{abstract}
As Artificial Intelligence (AI) systems proliferate, the need for systematic, transparent, and actionable processes for evaluating them is growing. 
While many resources exist to support AI evaluation, they have several limitations.
Few address both the \emph{process} of designing, developing, and deploying an AI system and the \emph{outcomes} it produces.
Furthermore, few are \emph{end-to-end} and  \emph{operational}, give \emph{actionable} guidance, or present evidence of usability or effectiveness in practice. 
In this paper, we introduce a third-party AI assurance framework that addresses these gaps.
We focus on \emph{third-party} assurance to prevent conflict of interest and ensure credibility and accountability of the process.
We begin by distinguishing assurance from audits in several key dimensions.
Then, following design principles, we reflect on the shortcomings of existing resources to identify a set of \emph{design requirements} for AI assurance.
We then construct a \emph{prototype} of an assurance process that
consists of (1) a responsibility assignment matrix to determine the different levels of involvement each stakeholder has at each stage of the AI lifecycle, (2) an interview protocol for each stakeholder of an AI system, (3) a maturity matrix to assess AI systems' adherence to best practices, and (4) a template for an assurance report that draws from more mature assurance practices in business accounting.
We conduct \emph{early validation} of our AI assurance framework by applying the framework to two distinct AI use cases---a business document tagging tool for downstream processing in a large private firm, and a housing resource allocation tool in a public agency---and conducting six expert validation interviews.
Our findings show early evidence that our AI assurance framework is sound and comprehensive, \emph{usable} across different organizational contexts, and \emph{effective} at identifying bespoke issues with AI systems. 
\end{abstract}

\ccsdesc[500]{Software and its engineering~Software development process management}
\ccsdesc[500]{Social and professional topics~Quality assurance}
\ccsdesc[500]{Social and professional topics~Technology audits}

\keywords{AI assurance, AI governance, AI evaluation, AI auditing}

\maketitle

\section{Introduction}

As Artificial Intelligence (AI) systems are increasingly embedded in societal systems, there is a need for design, development, deployment, and evaluation mechanisms to assess whether AI systems are built in a way that is trustworthy, produce desired societal outcomes, and align with desired organizational and societal goals. However, many of the existing approaches to evaluating AI systems fall short in practice. 
Few look at both the \emph{process} by which the AI system was built as well as the \emph{outcomes} that resulted from each stage~\cite{mokander2023auditing} and most focus narrowly on technical stakeholders~\cite{kuehnert2025who, kawakami2024responsible, costanza2022audits, schiff2024emergence}. 
Many approaches are often too abstract to operationalize and put into practice~\cite{morley2019from, schiff2020principles, barletta2023rapid}
and lack concrete standards against which AI systems should be evaluated and improved~\cite{costanza2022audits}.
Moreover, the majority of existing approaches lack empirical validations demonstrating the approach's \emph{usability} and \emph{effectiveness}\footnote{
We define \emph{usability} as the extent to which the approach is easy to use by target users, and \emph{effectiveness} as the extent to which a causal relationship can be established between the use of the approach and its desired effects (for example, diagnosing problems with an AI system)~\cite{cartwright2009thing}.}~\cite{berman2024scoping}.
In line with the shortcomings identified above, we argue that there is a critical need for an AI evaluation approach that satisfies the following \emph{design requirements}: 
(R1) focuses on both the process and outcomes, (R2) addresses end-to-end AI system stages and corresponding stakeholders, (R3) offers clear instructions and guidance to conduct assessments, (R4) offers concrete recommendations for improvement, and (R5) is validated through empirical evidence of usability and effectiveness.

\begin{figure*}
    \centering
    \includegraphics[width=0.75\linewidth]{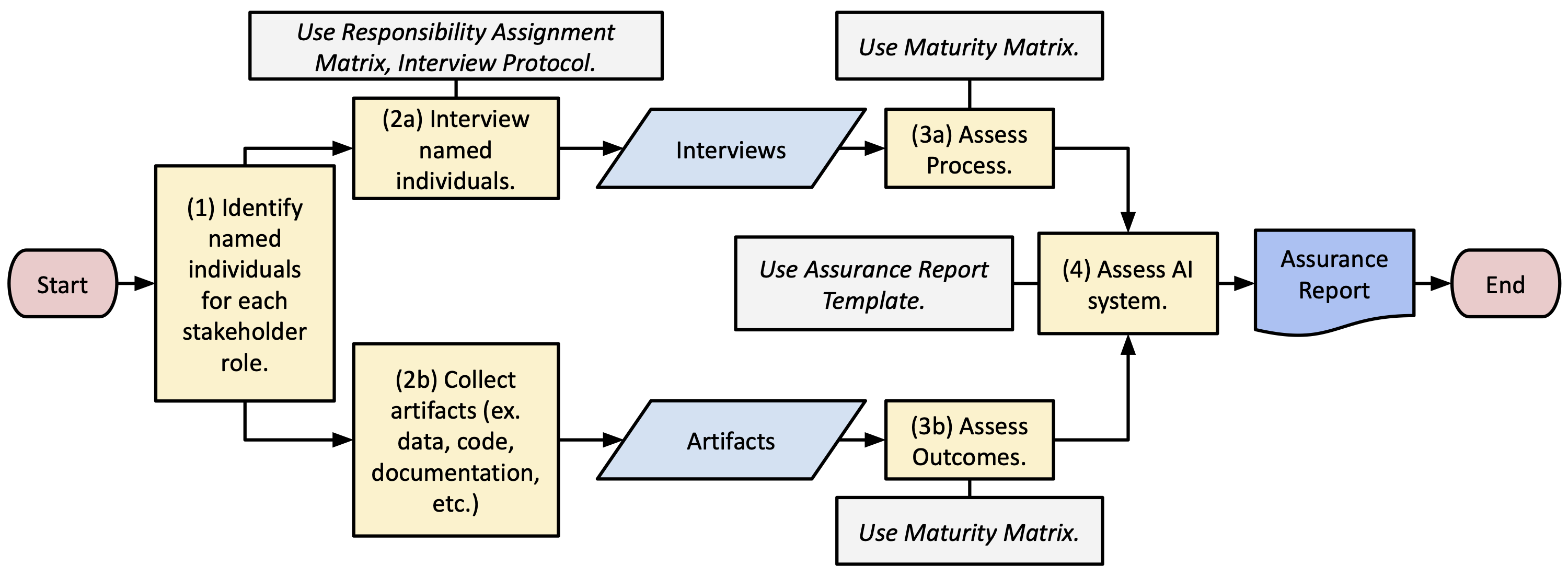}
    \caption{A flowchart of our AI assurance process. The figure specifies the resources that should be used in each step of the assurance process. The numbers correspond to the steps we describe in the main body of the paper (Section~\ref{sec:prototype}).}
    \label{fig:assurance-process}
\end{figure*}

In this work, we design a third-party\footnote{By third-party, we mean that the assessment is conducted by individuals external to the team designing, developing, and deploying the AI system.} 
AI assurance framework that addresses these design requirements (see an overview of the framework in Figure~\ref{fig:assurance-process})\footnote{The framework and materials can also be accessed at: \url{https://dssg.github.io/aiassurance/}.}.
We use \emph{assurance} to describe our evaluation approach, as the term naturally encompasses many of the previously-identified gaps in existing evaluation methods.
Borrowing from business accounting terminology, assurance describes an evaluation that focuses on both the outcomes and the \emph{processes} by which these outcomes are created, can be conducted throughout AI development, and is intended to be constructive.
While some prior work~\citep[e.g.][]{birhane2024ai, raji2020closing} has used the term AI \emph{auditing} to encompass a similar approach to evaluation, we choose to distinguish AI \emph{assurance} from \emph{auditing} to foreground a more \emph{proactive}, \emph{improvement}-oriented, and potentially \emph{voluntary} posture, in contrast to the more \emph{retrospective}, \emph{punitive}, and \emph{compliance}-triggered connotations generally associated with audits.
We focus on \emph{third-party}
assurance to minimize conflicts of interest, making it more likely to be impartial and trustworthy than an internal assurance~\cite{raji2022outsider, mokander2023auditing}.
Our framework is intended to be used by experts in AI assurance, auditing, or governance to evaluate an AI system at multiple points in its lifecycle. 

Guided by well-established principles of design~\cite{dam2025five, orr2024engineering, floyd1984systematic}, we develop a blueprint for AI assurance by following the three fundamental stages of design. 
First, we identify five \emph{design requirements} for an AI assurance process (Section~\ref{sec:requirements}). Second, we create a \emph{prototype} of an AI assurance framework that satisfies these design requirements (Section~\ref{sec:prototype}), containing four key components.
\begin{itemize}
    \item \textbf{Responsibility Assignment Matrix:} a matrix delineating the different levels and types of responsibility that each stakeholder has at each stage of the AI lifecycle (Appendix Table~\ref{tab:raci})
    \item \textbf{Interview Protocol:} a set of questions for each stakeholder to assess whether they discharge(d) their responsibilities appropriately (Appendix Section~\ref{sec:app-stakeholder-interview-protocol}).
    \item \textbf{Maturity Matrix:} a rubric that outlines best practices in every stage of the AI lifecycle, and assesses the maturity of the \emph{process} followed for designing, developing, and deploying an AI system, as well as the quality of the resulting \emph{outcomes} (Appendix Section~\ref{sec:app-maturity-matrix}).
    \item \textbf{Assurance Report Template:} a template for structuring the output of the assurance process, including recommendations for improvements to the system (Appendix Section~\ref{sec:app-report-format}).
\end{itemize}

Third, we perform \emph{early validation} of our assurance framework (Section~\ref{sec:testing}) in two ways.
We begin by applying the assurance framework to two real-world AI systems: a business document tagging tool for downstream processing in a large private company and a housing assistance resource allocation tool in a government agency.
These two cases differ across many dimensions, including context, scale, and stakeholder involvement. 
Next, we conduct expert validation interviews with individuals who are currently industry practitioners or have extensive practitioner experience in AI assurance, auditing, or governance.
We conduct these validations to gather initial evidence on our framework's necessity, soundness, usability, and effectiveness. 
The two validation methods complement each other: the research team's pilot applications of the framework offer deep insights into the experience of applying the framework and the specific issues it can surface in practice, while expert validation interviews offer a shallower, but external, perspective assessing the framework’s perceived usability and effectiveness.
Our findings provide initial positive evidence that our framework is necessary, sound, usable, and effective. However, additional evidence is necessary for comprehensive validation of our framework, and our early assessment identifies important practical considerations that future versions of the framework must account for, including the need for flexibility in conducting assurance and the importance of assurance timing.

The primary contribution of this work is a blueprint for third-party assurance of AI systems, accompanied by early evidence demonstrating its necessity, soundness, usability, and effectiveness. While further assessment—particularly longitudinal and large-scale studies—is required to fully establish the practical value and long-term impact of this approach, the proposed framework provides a concrete and well-founded starting point for the development of standardized AI assurance procedures. In more mature industries, assurance practices are supported by established standards and carried out by trained, independent professionals. We envision a similar trajectory for AI, in which qualified assurance professionals can reliably evaluate AI systems across organizational contexts. Such standardization is especially critical given that many organizations developing or deploying AI systems lack the capacity or resources to design and implement bespoke assurance frameworks. By articulating a structured, end-to-end approach to third-party AI assurance, this work represents an important step toward that future.
\section{Background \& Related Work}\label{sec:related-work}

\xhdr{Responsible AI (RAI) Tools}
There is a growing body of tools created to guide the ethical design, development, and deployment of AI systems, such as Datasheets for Datasets~\cite{gebru2021datasheets}, Data Statements~\cite{bender2018data}, and Model Cards~\cite{mitchell2019model}.
However, RAI has continued to be proven difficult to operationalize, for several reasons: there is an outsized focus on the technical stages and stakeholders~\cite{kaye2023risky, kuehnert2025who}, frameworks are often too high-level~\cite{morley2019from, schiff2020principles, barletta2023rapid} or too narrow~\cite{kaye2023risky}, and they are often difficult to apply to real-world use cases~\cite{ojewale2025towards, groves2024auditing}.
In this paper, we create an AI assurance framework that addresses several of these difficulties.

\xhdr{Internal AI Governance and Risk Management}
Standards and regulatory bodies have produced guidance for AI governance and risk management, such as the NIST AI Risk Management Framework~\cite{nist2023artificial}, SR 11-7 Guidance on Model Risk Management~\cite{board2011sr117}, and ISO/IEC 42001~\cite{iso2023iso}.
In academic literature,~\citet{rismani2023plane} find that frameworks from other disciplines, such as safety engineering, can be applicable for AI risk management. 
Algorithmic impact assessments~\cite{canada2024algorithmic, mantelero2018ai, reisman2018algorithmic} have also been proposed as tools for governance throughout the AI lifecycle.
However, the previously mentioned guidance, frameworks, and tools are all intended primarily for organizations to use \emph{internally} to govern the AI systems they develop and use.
Our focus is on \emph{external, third-party assurance} as a mechanism to establish stronger trustworthiness than internal processes can provide~\cite{raji2022outsider, mokander2023auditing, ajunwa2020auditing}.

\xhdr{AI Audits}
AI assurance is closely related to, but distinct from, AI audits. 
AI audits ``evaluate [AI systems] according to a specific set of criteria and provide findings and recommendations to the auditee, to the public, and/or to another actor, such as to a regulatory agency or as evidence in a legal proceeding''~\cite{costanza2022audits}.
AI audits can be performed internally by the company creating the system, or externally by a third party. 
There are tradeoffs to both approaches, such as a potential lack of access to critical data and information in a third-party audit, and a potential for conflicts of interest for internal audits~\cite{raji2022outsider, costanza2022audits}.
As in other industries, AI audits should compare behavior with a predefined set of standards, include concrete post-audit plans (e.g. disclosure), and be conducted by auditors who are trained and accredited through standardized processes (although such processes are nascent)~\cite{lam2024framework, raji2022outsider, costanza2022audits, mokander2023auditing, mokander2021ethics}.

While there are many tools that can assist in certain aspects of AI audits, such as questionnaires to audit datasets~\cite{gebru2021datasheets} and models~\cite{mitchell2019model}, there are fewer end-to-end frameworks.
However, there are some notable exceptions. 
~\citet{raji2020closing} and~\citet{lam2024framework} propose end-to-end frameworks: one for organizations to conduct internal auditing, and the other for auditing AI systems for compliance with New York City Local Law 144, respectively.
The implicit goal of AI audits is to hold organizations \emph{accountable} for their AI systems.
Unfortunately, accountability outcomes are rarely realized.
~\citet{birhane2024ai} outline several reasons for this, including an outsized focus on auditing the model outcomes, which has prevented auditors from understanding the AI system more holistically within its organizational structures.
Towards addressing this gap, we focus on AI \emph{assurance} (which includes organizational processes, among other differences) rather than audits.
We provide an initial attempt at defining and synthesizing a set of standards for AI systems, outlining an accessible report format through which assurance results can be communicated, and recommending critical elements that must be included in AI assurance training.
Our work integrates and operationalizes frameworks such as those set out by~\citet{raji2020closing} and~\citet{lam2024framework} through providing a concrete interview protocol, maturity matrix, and reporting format, and applying these artifacts to real-world rather than hypothetical AI systems.


\xhdr{Assurance vs. Audits in Other Industries}
In business accounting, an ``audit'' consists of the review of a business’s statements, reports, and records to verify the accuracy of the respective claims relative to the activities they represent. 
In contrast, an ``assurance'' is the examination of the \emph{processes} by which the records are produced~\cite{lewis2023audit}. 
In addition to reviewing a business's statements (that is, conducting an audit), assurances involve gathering employee feedback, observing organizational behavior, and looking at quality control mechanisms~\cite{lewis2023audit}.

We make a similar distinction between \emph{AI} audits and assurance~\cite{uk2024introduction, us2024overview, chalmers2021trust}.
Specifically, AI audits and assurance differ across three key dimensions: first, audits focus on the \emph{outcomes} of an AI system, while assurance focuses both on the outcomes and the processes by which the outcomes are created. 
Second, audits are primarily conducted \emph{after} a model has been deployed, whereas assurances are conducted \emph{throughout} the AI lifecycle. 
Third, audits are more punitive, determining what is wrong with an AI system, and how it fails to comply with applicable norms and regulations. Assurance, on the other hand, is often more constructive: it involves continuously working with stakeholders to mitigate risks and ensure the trustworthiness of the AI system.
Importantly, audits can be thought of as one (albeit major) element of assurance~\cite{uk2024introduction}.
Formally, we adopt the definition of AI assurance as ``the \emph{techniques}, \emph{activities}, and \emph{processes} used to evaluate and assure expected properties of AI components or AI-enabled systems \emph{throughout the lifecycle} of these components or systems''~\cite{us2024overview}.

We note that prior work has sometimes used ``audits'' to encompass aspects of what we define to be ``assurance''.
For example,~\citet{birhane2024ai} use the term ``ecosystem audits'' to include the processes involved in designing, developing, and deploying AI systems, and~\citet{raji2020closing} frame audits as end-to-end evaluations conducted throughout the AI lifecycle.
Other work has distinguished between ``process audits'' and ``technology-oriented audits''~\cite{mokander2023operationalising, mokander2023auditing} (we use the term \emph{assurance} to refer to the combination of process and technology-oriented audits, and the term \emph{audit} to refer to technology-oriented audits). 
In this paper, we deliberately choose to use the term ``assurance'' to foreground a \emph{proactive}, \emph{improvement}-oriented, and potentially \emph{voluntary} posture, in contrast to the more \emph{retrospective}, \emph{punitive}, and \emph{compliance}-triggered connotations often associated with the term ``audit''.
This distinction reflects a normative shift towards treating system evaluation as an ongoing process of iterative improvement rather than a series of episodic inspections triggered by external scrutiny or failure---a potential mechanism to contend with some of the challenges with AI auditing.


\section{Our Proposed Requirements for an AI Assurance Process}\label{sec:requirements} 

In this paper, we propose a framework that enables third-party assurance professionals to conduct AI assurance across a range of AI systems\footnote{
We adopt the Organization for Economic Co-operation and Development (OECD)'s definition of an AI system as ``a machine-based system that, for explicit or implicit objectives, infers, from the input it receives, how to generate outputs such as predictions, content, recommendations, or decisions that can influence physical or virtual environments. Different AI systems vary in their levels of autonomy and adaptiveness after deployment''~\cite{russell2023updates}.}  and organizational contexts, with the goal of standardizing how AI systems are assessed.
Analogous to financial assurance, the objective of AI assurance is to establish trust between the public and the organizations that develop or use AI systems; to enable early detection of risks; and to support compliance with standards, best practices, and regulatory requirements. The design of our framework is explicitly oriented toward achieving these goals.

By enabling independent, third-party use, the framework can mitigate conflicts of interest and promote a more objective, credible assurance process, ultimately increasing confidence in its outcomes. While we focus on third-party use, the framework is equally applicable to qualified first and second parties seeking to assess and improve AI systems.
Additionally, the framework is designed to be applied at multiple stages of the AI lifecycle, enabling the identification of potential issues earlier, when they are typically easier and less costly to address.
Furthermore, the framework provides a set of materials that serve both as assessment criteria for assurance professionals and as a reference for organizations developing or deploying AI systems. While ad hoc audits or one-off assessments
may surface issues similar to those identified through our framework, such efforts lack the broader benefits of a standardized approach, including consistent expectations, comparability across systems, and the establishment of shared best practices across organizations.

Here, we develop five \emph{design requirements} for such a framework based on the existing gaps in the current space of RAI tools.
For each, we identify how the requirement has been insufficiently addressed in the status quo, and our contribution towards filling the gap.

\xhdr{R1: Focuses on both process and outcome}
\citet{mokander2023auditing} outlines two types of auditing: \emph{process audits}, which focus on technology providers' governance structures and quality management systems, and \emph{technology-oriented audits}, which focus on the properties and capabilities of AI systems.
He highlights that while research on both process audits and technology-oriented audits is flourishing, there is little dialogue between the two strands of research.
He further argues that linking of the two types of audits is necessary for an effective AI audit.

Our AI assurance framework is an initial attempt to do so: we focus on both the \emph{process} that was followed when designing, developing, and deploying the AI system through interviewing the system's stakeholders, and the \emph{outcomes} that came out of the process through analyzing artifacts such as data, models, and experimental results.
We note that our framework does not attempt to establish a causal relationship between process quality and system outcomes. 
Rather, we rely on the widely accepted premise that system design and governance processes shape downstream behavior, while recognizing that rigorous causal validation would require longitudinal, cross-domain empirical study.
Our framework evaluates both process and outcomes because each provides complementary but incomplete signals on system quality. 



\xhdr{R2: End-to-end---across AI system stakeholders and AI lifecycle stages}
Previous work identifies that many RAI tools focus narrowly on a core set of technical stages (e.g. statistical modeling) while neglecting other highly influential stages (e.g. value proposition)~\cite{kaye2023risky, kuehnert2025who}.
~\citet{kawakami2024responsible} found that existing RAI tools often treat end-users as an afterthought.
There have been calls to create auditing processes that focus on both technical and qualitative analyses and encourage more robust stakeholder involvement~\cite{costanza2022audits, schiff2024emergence}.

Our proposed AI assurance framework addresses all of the key AI system stakeholders---leaders, designers, developers, deployers, end-users, and impacted communities---and all the AI lifecycle stages from value proposition to monitoring. 
Our end-to-end approach allows for a holistic assessment of AI governance practices across an entire AI system, broadening the perspective from the largely technical one present today~\cite{ojewale2025towards, black2023toward}. 

\xhdr{R3: Offers clear instructions and guidance to conduct assurance} 
While many frameworks offer aspirational principles, they often remain too high-level to be operationalized~\cite{morley2019from, schiff2020principles, barletta2023rapid, mandal2025governing}, or remain difficult for some stakeholders, especially non-technical ones, to use and understand~\cite{kuehnert2025who, flanagan2025methodology}. 
However, an AI audit or assurance process must be adequately structured to be effective~\cite{mokander2023auditing}.
This gap highlights the need for an assurance framework that offers clear instructions to conduct assurance.

Every component of our proposed AI assurance framework 
provides concrete guidance for the assurance professional.
The interview protocol contains a set of questions to ask each stakeholder. 
The maturity matrix assesses an AI system against best practices.
Each row of the maturity matrix outlines a property that an AI system should be tested for, and clearly defines the criteria for different levels of maturity.
This structure of the maturity matrix ensures that AI systems can be evaluated and understood by practitioners and organizational leaders with minimal ambiguity. 

\xhdr{R4: Offers concrete recommendations for improvement} 
AI audits, and more broadly AI assurance, evaluate an AI system against a specific set of criteria~\cite{costanza2022audits} and should present metrics that clearly quantify the problematic behaviors of the AI system~\cite{bandy2021problematic}. However,~\citet{costanza2022audits} found from a field scan of the AI audit ecosystem that the standards available for evaluating AI systems are, at best, minimal.
A more recent study by~\citet{schiff2024emergence} found an emerging convergence around the standards set by the EU AI Act~\shortcite{eu2024ai} and the U.S. NIST AI Risk Management Framework~\shortcite{nist2023artificial}.
However, these standards remain quite broad.

In our proposed AI assurance framework, informed by analyzing previous academic literature and developing real-world AI systems, we define what best practices would entail in every stage of the AI lifecycle.
We further divide each stage of the AI lifecycle into ``subcategories'' and define best practices within each subcategory.
By doing so, we implicitly provide concrete guidance for improvement, delineating clear steps for organizations to improve their practices.

\xhdr{R5: Contains validations demonstrating evidence of usability and effectiveness}
~\citet{kuehnert2025who} found that approximately two-thirds of existing RAI tools did not show even \emph{suggestive} empirical evidence of usability or effectiveness in practice. 
However, previous work has highlighted the importance of the context of use when conducting AI audits or, more broadly, using RAI tools~\cite{rismani2023plane, kijewski2024rise} and encouraged researchers to conduct empirical studies on RAI tools' usability, feasibility, and effectiveness~\cite{berman2024scoping, mokander2023auditing, kuehnert2025who}.
~\citet{flanagan2025methodology} call for the partnership of industry and academia in the creation of AI evaluations.

We address the importance of understanding context of use and conducting empirical studies on RAI tools by piloting our proposed AI assurance framework on two \emph{real-world} AI systems and conducting expert validation interviews.
Each AI system in our pilot applications has distinct purposes and is designed, developed, and deployed in different organizational contexts.
We show initial evidence of our framework's \emph{usability} (the extent to which our assurance framework is applicable and easy to use) and \emph{effectiveness} (the extent to which our assurance framework can diagnose problems within an AI system, suggest appropriate solutions, and improve the practices surrounding the AI system~\cite{berman2024scoping}).
By reporting our findings, we emphasize the importance of validating RAI tools with multiple methodologies.

\section{Our Proposed Prototype for an AI Assurance Framework}\label{sec:prototype}

Now, we provide a \emph{prototype} of an assurance framework that satisfies our design requirements.
Figure~\ref{fig:assurance-process} contains a high-level schematic.
We discuss the framework's steps and conclude with guidance on intended use.

\xhdr{Step 1: Identify named individuals for each stakeholder role} 
Based on the NIST AI Risk Management Framework~\shortcite{nist2023artificial} and the work of~\citet{kuehnert2025who}, we select a key set of stakeholders within an AI system: leaders, designers, developers, deployers, end-users, and impacted communities (Appendix Table~\ref{tab:stakeholders}). 
Leaders are responsible for setting the organization’s strategic goals and high-level policies, typically occupying senior or executive roles.
Designers translate business and policy needs into technical problems, envision user experience, and work across functions to align AI solutions with stakeholder needs.
Developers use technical expertise to curate data and train AI models to perform the specified tasks.
Deployers are responsible for piloting, validating, and integrating the AI system into existing workflows and ensuring a reliable user experience.
End-users interact with and apply the AI system’s output to make decisions or carry out organizational objectives.
Impacted community members include individuals or groups directly or indirectly affected by the AI system, along with their advocates and representatives.
The assurance professional should identify the individuals who correspond to each role for the AI system that they are conducting assurance on.
Note that depending on the AI system, there may be multiple people serving in each role or one person serving multiple roles.

\xhdr{Step 2a: Interview the named individuals} 
Each interview should ask questions related to each of the stages of the AI lifecycle: value proposition, problem formulation, data collection and processing, statistical modeling, testing and validation, deployment, and monitoring (Appendix Table~\ref{tab:stages}). 
The full interview protocol is provided in Appendix Section~\ref{sec:app-stakeholder-interview-protocol}.
We outline the definitions of the AI lifecycle stages here. Value proposition concerns determining the problem the AI system is designed to solve, the business and policy requirements, and whether using AI leads to a net-benefit over the status quo.
Problem formulation is about the translation of business needs and requirements to technical choices.
Data collection is about selecting, collecting, or compiling data to devleop and validate an AI model and how to sample, label, and link data.
Data processing is about making data usable by the model (e.g. cleaning, standardizing, or normalizing), including feature engineering.
Statistical modeling is about choosing the type of statistical model(s) on which to fit to the data, how to perform the fitting, model selection methodology, and metrics.
Testing and validation refers to the determination that the model is performing well enough for deployment.
Deployment is the process of deploying the model into a larger decision system.
Monitoring refers to how a model’s behavior is recorded and responded to over time to ensure there is no significant or unexpected degradation in performance and use over time.
We used previous work~\cite{black2023toward, nist2023artificial, ortega2024applying} to determine the key stages of the AI lifecycle.

We illustrate the different levels of responsibility that each stakeholder has in each stage of the AI lifecycle using a responsibility assignment matrix (Appendix Section~\ref{sec:app-maturity-matrix}).
Responsibility assignment matrices are widely-used project management tools that encourage an end-to-end view of a project and benefit organizations with multiple moving parts~\cite{rittenberg2024everything}.
The matrix breaks down a project into two dimensions: the involved stakeholders and the smaller stages within the project. 
Since different stakeholders do not have equal involvement in each stage of the AI lifecycle, we use a responsibility assignment matrix for AI systems to identify which stages of the AI lifecycle warrant further probing for each stakeholder.

\xhdr{Step 2b: Collect additional artifacts} Such artifacts should include data used in building the AI system, source code used (not just for developing the model but also for any analysis of experimental results), AI models developed, and any documentation for all the lifecycle stages.

\begin{table*}[]
    \centering
    \scriptsize
    \begin{tabular}{p{0.08\linewidth}p{0.195\linewidth}p{0.195\linewidth}p{0.195\linewidth}p{0.195\linewidth}}
        \toprule
         \textbf{Category} & \textbf{1---Lagging} & \textbf{2---Basic} & \textbf{3---Intermediate} & \textbf{4---Industry-Leading} \\
         \midrule
         \raggedright Organizational Feasibility: \emph{Process}
         & \raggedright The team has not considered how the task being solved is accomplished today, and how the AI system will be embedded into existing business processes. 
         & \raggedright The team has considered how the task being solved is accomplished today.
         The team has taken some steps to understand one aspect (best practices or business model). 
         There is room for improvement – for example, by getting approval/buy-in to how the AI system will either fit into existing business processes or require a change in business processes.
         & \raggedright The team has considered and has approval/buy-in to how the AI system will either fit into existing business processes or require a change in business processes.
         There is some room for improvement – the team has a good understanding of existing business processes, but this knowledge could be further supplemented by asking additional people or consulting more existing documentation, for example. 
         &  \raggedright The team has considered and has approval/buy-in to how the AI system will either fit into existing business processes or require a change in business processes. \tabularnewline
         \midrule
         \raggedright Organizational Feasibility: \emph{Outcome}
         & \raggedright The status quo is better than the AI system across all the relevant organizational feasibility dimensions.
         & \raggedright The AI system aligns with some of the existing best practices and somewhat fits into the business model.
         There are some major changes to workflows as a result of the AI system, and considering these changes, the AI system does not fit in well with the business model.
         & \raggedright The AI system aligns with most of the existing best practices and mostly fits into the business model. 
         There are some major changes to workflows as a result of the AI system, but considering these changes, overall, the AI system still fits into the business model.
         & \raggedright The AI system aligns with existing best practices and fits into the business model. \tabularnewline
         \bottomrule
    \end{tabular}
    \caption{Example rows of the maturity matrix. Organizational feasibility is a \emph{subcategory} of the broader AI lifecycle stage value proposition, and is about whether the AI system aligns with or requires changes in existing business workflows. The rows illustrate how we grade the \emph{process} (top) the \emph{outcomes} (bottom).}
    \label{tab:maturity-matrix-example}
\end{table*}

\xhdr{Step 3: Conduct an assessment of both the (a) \emph{process} and (b) \emph{outcomes} of the AI system using the maturity matrix}
The assessments should be conducted using the AI maturity matrix (Appendix Section~\ref{sec:app-maturity-matrix}) that we developed using previous research~\cite{kuehnert2025who, ojewale2025towards}, our experience, and industry best practices. 
Maturity matrices are commonly used to assess where an organization currently stands in terms of best practices and provide them with steps they can take to increase their maturity. 
We designed the AI maturity matrix to assess each stage of the AI lifecycle through a set of \emph{subcategories}.
These subcategories were determined by surveying past work, in particular the RAI tools used by~\citet{ojewale2025towards} and~\citet{kuehnert2025who}, and analyzing real-world AI systems.
Each subcategory is assessed on both the \emph{process} that was followed when designing, developing, and deploying the AI system and the \emph{outcomes} that resulted.
Each subcategory's process or outcomes can be assessed as Lagging, Basic, Intermediate, or Industry-Leading. 
We designed the subcategories with the intention of making them use-case agnostic and weighing them equally in the assurance process.
Table~\ref{tab:maturity-matrix-example} contains two example rows (process and outcome) for the same subcategory. 

\xhdr{Step 4: Review the assessments and produce the assurance report} 
Using the assessments from the maturity matrix, we grade the \emph{entire} AI system with one of the following using language from financial auditing and assurance~\cite{russo2023what}: 
\begin{enumerate}
    \item Inadequate access to perform the evaluation.
    \item Adequate access, lack of compliance with baseline industry expectations.
    \item Adequate access; adequate compliance with baseline expert expectations but lack of evidence of compliance with state-of-the-art and best practices.
    \item Adequate access, adequate compliance with best practices, and state-of-the-art industry standards.
\end{enumerate}
In these ratings, ``baseline industry expectations'' and ``state-of-the-art and best practices'' were determined by consulting previous literature (namely the RAI tools in~\citet{ojewale2025towards} and~\citet{kuehnert2025who}) and our team members' experiences designing, developing, and deploying AI systems in real-world organizations.
Our definitions for ``baseline expert expectations'' in each subcategory of each AI lifecycle stage are reflected in the ``Basic'' column of the maturity matrix; our definitions of for ``state-of-the-art and best practices'' are reflected in the 
``Intermediate'' and ``Industry-Leading'' columns of the maturity matrix.
Appendix Section~\ref{sec:app-report-format} provides more details, the rules for determining the final grade using the maturity matrix, and the structure for the final report.

\xhdr{Intended Use}
Similar to how financial audits and assurances are conducted by trained (and often certified) experts  and large firms in the area, we intend for the AI assurance framework to be used by third-party individuals and organizations who have expertise in AI assurance, auditing, or governance.
When applying the AI assurance framework, the assurance professional should also determine the relevant existing regulations and standards for the AI system (e.g. SR 11-7~\cite{board2011sr117}), check whether they are considered, and if the AI system complies.
Assurance should occur at multiple points throughout the AI lifecycle, ideally after each stage, though the exact frequency will depend on practical constraints, the organization, and the use case. Each engagement should produce recommendations grounded in the maturity matrix, drawing on the next-level criteria where the organization has not yet reached ``Industry-Leading''. In subsequent engagements, professionals can evaluate whether these recommendations have been implemented and, if not, identify barriers and issue revised or alternative guidance.
\section{Early Testing: Necessity, Soundness, Usability, and Effectiveness}\label{sec:testing}

We want to demonstrate that our AI assurance framework is \emph{usable} (easy to use) and \emph{effective} (can diagnose problems within an AI system, suggest appropriate solutions, and improve the practices surrounding the AI system)~\cite{berman2024scoping}.
A true assessment of \emph{usability} would entail applying the assurance framework to multiple different real-world AI systems, and assessing whether it applies to those systems and how easy it is to do so.
A true assessment of \emph{effectiveness} would entail a randomized longitudinal study that uses our framework to diagnose problems with an AI system and suggest appropriate solutions, works with the organization to implement the solutions, and conducts the assurance process over time to see whether the AI system has improved (compared to other approaches or the status quo at the organization). 

While these multi-year assessments of usability and effectiveness are important work, we consider them out-of-scope for the present contribution.
Rather, in this paper, we confirm the necessity of our framework, verify its soundness, and demonstrate \emph{initial} evidence of usability and effectiveness by conducting \emph{early testing} of our AI assurance framework in two distinct ways.
First, we apply our assurance framework to two real-world AI systems designed, developed, and deployed in different organizational settings.
The pilot applications of the AI assurance framework by two fairness, accountability, and transparency (FAccT) researchers on the research team provide first-person insights on the process of trying to conduct third-party assurance using the framework.
Furthermore, this validation methodology allows for deeper insights into specific AI systems and the types of problems and recommendations that our process is able to surface.
Second, we conduct expert validation interviews with current practitioners or individuals who have extensive practitioner experience in AI assurance, auditing, or governance.
The expert validation interviews act as a heuristic evaluation providing an external practitioner perspective on the anticipated usability and effectiveness of our process.
The two validation approaches are complementary: one provides in-depth, first-person insights, while the other offers higher-level, third-person perspectives.
Our validations are crucial, formative steps in the full evaluation of our framework. 
In an RAI landscape where two-thirds of tools provide no in-tool indication of empirical evidence pointing to its potential usability and effectiveness~\cite{kuehnert2025who}, we view our validations, while formative, as a critical, underexplored, endeavor.

\subsection{Methodology}

\xhdr{Pilot Applications}
To test and fine-tune our assurance framework, we used the following AI systems.
\begin{enumerate}
    \item \textbf{Business document tagging tool for downstream processing in a large private firm (with 1000+ employees)}. 
    We interviewed 4 employees working on an AI-enabled system designed to tag business documents with a standardized tagging taxonomy and increase the efficiency of downstream processing.
    Together, they encompassed the roles: leaders, designers, developers, and deployers (Table~\ref{tab:stakeholder-participants}). At the time of assurance, this AI system had been decommissioned.
    \item \textbf{Housing assistance resource allocation in a public-sector agency (with 1000+ employees)}. We interviewed 7 individuals involved in an AI-augmented process intended to improve the existing housing resource allocation process.
    Together, they encompassed the roles: leaders, designers, developers, deployers, and end-users (Table~\ref{tab:stakeholder-participants}). 
    At the time of assurance, this AI system was in testing and validation.
\end{enumerate}

In line with the iterative nature of the design process~\cite{dam2025five, orr2024engineering}, we use our pilot applications as a feedback mechanism to refine our framework.
We began by conducting interviews with stakeholders of the first AI system, collecting insights that helped us identify recurring gaps in existing practices. 
These insights, along with the RAI tools in the systematic reviews by~\citet{ojewale2025towards} and~\citet{kuehnert2025who}, informed the initial design of the maturity matrix. 
We applied our maturity matrix to the first AI system, which allowed us to identify gaps in the interview protocol; specifically, gaps in questions for certain roles and lifecycle stages.
We revised our protocol to address these gaps before moving on to the second AI system.  
By conducting the interviews for the second AI system and using our maturity matrix, we further validated the usability and effectiveness of our assurance framework. 
Each interview lasted 1-2 hours.
The interview protocol is in Appendix Section~\ref{sec:app-stakeholder-interview-protocol} and was approved by an Institutional Review Board (IRB).

\begin{table}[]
    \scriptsize
    \centering
    \begin{subtable}[t]{0.48\textwidth}
        \centering
        \caption{Stakeholder Participants}
        \label{tab:stakeholder-participants}
        \begin{tabular}{cp{0.4\textwidth}p{0.4\textwidth}}
             \toprule
             \textbf{ID} & \textbf{AI System} & \textbf{Stakeholder Role} \\
             \midrule
             S1 & \raggedright Business Document Tagging Tool & Designer \\
             S2 & \raggedright Business Document Tagging Tool & Developer \\
             S3 & \raggedright Business Document Tagging Tool & Developer, Deployer \\
             S4 & \raggedright Business Document Tagging Tool & Leader \\
             S5 & Housing Resource Allocation & Designer, Developer, Deployer \\
             S6 & Housing Resource Allocation & Designer, Developer, Deployer \\
             S7 & Housing Resource Allocation & Designer, Developer, Deployer \\
             S8 & Housing Resource Allocation & Leader \\
             S9 & Housing Resource Allocation & End-user \\
             S10 & Housing Resource Allocation & End-user \\
             S11 & Housing Resource Allocation & End-user \\
             \bottomrule
        \end{tabular}
    \end{subtable}
    \hfill
    \begin{subtable}[t]{0.48\textwidth}
        \centering
        \caption{Expert Participants}
        \label{tab:expert-participants}
        \begin{tabular}{cp{0.29\textwidth}p{0.2\textwidth}p{0.29\textwidth}}
             \toprule
             \textbf{ID} & \textbf{Organization Type} & \textbf{Role} & \textbf{Expertise} \\
             \midrule
             E1 & \raggedright Large Financial Services Company & \raggedright Data Scientist & Assurance, Auditing, Governance \\
             E2 & \raggedright Large Financial Services Company & \raggedright Director & Auditing, Governance \\
             E3 & \raggedright Non-Profit & Director & Assurance, Auditing, Governance \\
             E4 & \raggedright Large Automotive Corporation & \raggedright Counsel & Assurance, Auditing, Governance \\
             E5 & \raggedright Large Automotive Corporation & \raggedright Counsel & Assurance, Auditing, Governance \\
             E6 & \raggedright Large Financial Services Company & \raggedright Director (Former) & Assurance, Auditing, Governance* \\
             \bottomrule
        \end{tabular}
    \end{subtable}
    \caption{Interview participants in (a) our pilot applications of the AI assurance framework and (b) expert validation interviews. Some participants played multiple stakeholder roles within the relevant AI system. We refer to participants in (a) and (b) as ``stakeholders'' and ``experts'' in our findings, respectively. Note (*): E6 stated that they had moderate expertise in each of AI assurance, auditing, and governance.}
\end{table}


\xhdr{Expert Validation} 
We conducted semi-structured interviews with six experts in AI assurance, auditing, or governance (Table~\ref{tab:expert-participants}).
These interviews intend to serve as a heuristic evaluation of our framework.
We began by asking experts about their background as it relates to AI assurance, auditing, and governance. 
Then, we asked about current evaluation and assurance practices for AI systems within their organization and field.
We provided a 10-minute presentation of our framework, then asked about their awareness of any similar third-party frameworks, our framework's soundness (e.g. whether the framework has any major gaps), and anticipated usability\footnote{
When constructing the questions, we consulted typical considerations in usability research, such as the aspects highlighted by the System Usability Scale~\cite{brooke1996sus},~\citet{nielsen1994usability}, and~\citet{olsen2007evaluating}; however, we choose only certain aspects from these sources in line with the properties that are most critical for an AI assurance process as it is important to map evaluations to the claims we wish to make about the framework~\cite{ledo2018evaluation, greenberg2008usability}.}
and effectiveness\footnote{
RAI effectiveness evaluation is still nascent area~\cite{berman2024scoping}, meaning that there are not agreed upon aspects of effectiveness. Thus, we focus our questions on the broad goals of AI assurance which are improved organizational practices and AI system outcomes.}.
Interviews lasted 1-1.5 hours; see the IRB-approved protocol in Appendix Section~\ref{sec:app-expert-interview-protocol}.

We recruited experts using convenience and snowball sampling, a common approach for hard-to-reach populations such as AI assurance experts~\cite{heckathorn2011comment, lee2021landscape}.
We first identified individuals in the research team's professional network and industry organizations that conduct AI assurance, audit, or governance work.
We reached out to these individuals and organizations via email to initialize a snowball sample, inquiring whether they would be open to participating or forwarding the email to any other people that might be a good fit.


\xhdr{Data Analysis} 
For both the stakeholder and expert interview transcripts, we use an inductive thematic analysis approach~\cite{braun2006using}, allowing themes to emerge from the data. 
This approach enables us to identify cross–use case issues in AI systems, synthesize the framework’s strengths and shortcomings, and surface any unexpected considerations from the data.
Across the two types of interviews, 2 of the co-authors coded 3 transcripts separately.
Upon discussion, they found that there was substantial agreement in the codes and themes.
When disagreements occurred, they were primarily around wording (e.g. ``scaling issues'' vs. ``inefficiency and resource demand of assurance'').
Thus, the remaining transcripts were coded by 1 co-author.
The coders attended most interviews simultaneously and continued to meet regularly to arrive at a shared understanding of the findings.

\xhdr{Scope of Validations}
We emphasize that our evaluation provides \emph{exploratory} rather than \emph{confirmatory} evidence of effectiveness. 
Our aim was to assess whether the framework is \emph{usable in practice} and \emph{capable of surfacing meaningful issues} rather than to establish any causal claims.
These pilot applications are also limited in that both are retrospective examinations of existing systems, one of which had already been decommissioned.
As such, our validations do not provide conclusive evidence about the overall effectiveness about the assurance process.

\subsection{Findings}

We organize our findings from our pilot applications of the AI assurance framework and expert validation interviews into four sections: (1) the necessity for a third-party AI assurance framework like ours, and its (2) soundness, (3) usability, and (4) effectiveness.
We use \emph{stakeholders} to refer to participants in our pilot applications, \emph{experts} to refer to participants in the expert validation interviews, and \emph{participants} to refer to the stakeholders and experts together.

\xhdr{Necessity: Participants pointed to the need for a standardized third-party assurance process}
When asked about current practices in AI assurance, auditing, evaluation, and governance, 
both stakeholders and experts highlighted a lack of standards: governance was internal, ad-hoc, and use-case specific.
Some organizations did not have organizational standards for AI systems (e.g. \textit{``no central team [...] dictating [...] documentation standards''} (S4)); as a result, standards were defined on an ad-hoc and use case-specific basis by individual team leads.
Others had organization-wide standards, although they were internal.
In particular, when asked for any other end-to-end frameworks that help people conduct assurance, most experts pointed to \emph{internal} (proprietary) frameworks, \emph{internal} governance standards~\citep[e.g.][]{board2011sr117}, or third-party tools supporting \emph{internal} processes (e.g. Credo AI\footnote{https://www.credo.ai/}).
E4 stated that companies like McKinsey had a process similar to ours; however, we could not find publicly available information on this process.
These findings are consistent with prior academic work showing an outsized emphasis on internal governance mechanisms~\cite{raji2022outsider}.

Consequently, we find that a third-party AI assurance framework fills an important gap in the current landscape and is therefore \emph{necessary}: current standards are inconsistent and largely private, leaving no consistent standards for AI assurance that are applicable across organizations.
E1 said our framework was \textit{``very helpful''} because \textit{``it promotes [...] a mutual understanding of what good looks like''}.
Our framework provides a standardized approach to assurance where no prior publicly available frameworks exist~\cite{costanza2022audits, schiff2024emergence}.

\xhdr{Soundness: Participants perceived our framework as comprehensive}
Most experts found that writ large, our AI assurance framework is sound: the framework is truly end-to-end across both stakeholders and stages, the definitions of the stakeholders and stages are reflective of real-world systems, the separation between process and outcome is understandable and reasonable, and the descriptions of the maturity matrix grades are reflective of today's practices.
E3 particularly appreciated the parallel structure and the inclusion of both process and outcomes, highlighting that typically \textit{``people fight over whether [assurance is] supposed to look at the process or the outcomes''} or \textit{``focus on one or the other''}.

While our stakeholder and stage definitions are perceived as accurate, participants also highlighted several potential improvements.
E1 stated that the definitions for designers and deployers could be too broad for some contexts.
In contrast, we found from our pilot applications of the framework that despite the apparent broadness of our definitions, stakeholders reported taking on multiple roles throughout the project (S3, S5, S6, S7), complicating attempts to draw sharp boundaries between responsibilities.
Moreover, participants highlighted that the stages of AI development are often more iterative 
than most theoretical structures assume. 
E2 and stakeholders from both use cases (S3, S6) noted that while it is useful to have AI systems broken down into the stages in Appendix Table~\ref{tab:stages}, AI development often blends phases or revisits prior stages.
Additionally, experts (E3, E4, E5) highlighted changes that could be made to stakeholders and stages for cases where companies use third-party AI models, such as the addition of ``integrators'' or skipping and replacing certain stages of the AI lifecycle.

\xhdr{Usability: Participants anticipated our framework to be usable in practice, but identified areas of improvement}
Overall, participants thought that our AI assurance framework was usable in practice. 
They especially noted its accessibility and learnability.
Experts saw the framework as \textit{``definitely in scope''} (E3) and anticipated there would be \textit{``no issues''} (E6) for professionals experienced in AI assurance, auditing, and governance to apply and believed they could \textit{``easily learn all [the parts of the framework]''} (E2).
At large, participants reflected that the structure of the framework felt familiar and resembled present-day \emph{internal} governance in mature organizations, making it accessible, easy to learn, and minimally error-prone.
This familiarity distinguishes our framework from others (e.g. safety engineering frameworks~\cite{rismani2023plane}).

Experts expressed mixed views on the likelihood and severity of errors in using the framework.
Some experts (E1, E2) did not view the process as prone to errors, and felt that even when errors were made, recovery from errors would \textit{``not [be] an issue''} (E2) due to the expertise of AI assurance professionals.
Other experts (E3, E4, E5) found the framework \textit{``clear enough''} (E3), but believed that errors could arise from assurance professionals’ skill-sets, inconsistencies across professionals’ judgments, and whether information needs are known before conducting the assurance.
They also cautioned that errors could be \textit{``quite severe''} (E3), depending on the risk of the AI system, and highlighted the need for clear instructions on how to apply the process.

Participants identified several considerations for the applicability of the framework. 
In our pilot applications, some stakeholders, especially those with a non-technical background, reported struggling with certain questions in our assurance framework.
S9 said,\textit{``I theoretically get [the assurance framework], but it's so far [...] outside the scope of [...] how I'm used to thinking about things that I don't know if I was helpful.''}
They also said that from their first-hand experience with the impacted community members (people in need of housing resources), impacted community members may not \textit{``understand [the interview questions] at all.''}
We further found that assurance processes must be able to accommodate specificity in definitions and procedures: varying understandings of terms like harm, data, AI system, and norms in different domains (e.g. finance vs. healthcare) may complicate consistent application. In fact, E6 commented several times that the framework was too high-level, while E3 pointed out that the framework was less applicable for general-purpose AI systems.
Moreover, E2 said that while the current version of the assurance framework accurately encompasses today's best practices, the AI development landscape is ever-changing; thus, it is critical to have a plan to update the framework when needed.

Finally, participants expressed that usability could be constrained by companies' incentives to adopt third-party assurance.
E4 noted that they could not \textit{``fathom any organization that I worked with [...] subjecting themselves to third-party assurance''} given that internal governance teams are often more familiar with both the business context and the specific AI system.
Experts (E4, E5, E6) suggested that creating a domain-specific process could increase companies' willingness to engage with third-party assurance.
Participants also noted the general monetary and temporal resource demands of assurance.
While governance processes were generally viewed as difficult to scale, participants suggested underscoring the framework's use in high-risk scenarios, asking less questions to stakeholders about stages they are less familiar with, and exploring the option to substitute interviews with emails as potential methods for reducing the practical constraints surrounding assurance. 

\xhdr{Effectiveness: Participants anticipated the framework to improve organizational processes and outcomes surrounding AI}
Overall, we find initial evidence of our framework's effectiveness through its potential to surface problems, motivate organizations to improve their practices, and find issues earlier in the AI lifecycle, where they are usually easier to address. 
From our pilot applications, we found that our assurance framework was effective at surfacing problems end-to-end across both the stakeholders and stages of the AI lifecycle; we provide a sample here.
We found that the business document tagging model was built as a result of the\textit{``aggressiveness of [the] business to show that they're doing something in the AI domain''} (S4).
This top-down push from leaders occurred despite the evidence surrounding the project's technical feasibility being scant.
During the testing and validation phase of the housing resource allocation project, S9, an end-user, expressed that there were no ways to obtain concrete information on the downstream impact of the model, despite wanting it.
While the problems we report here corroborate results found in existing research~\citep[e.g.,][]{heger2022understanding, godbole2025ai, faggella2018enterprise}, our framework's ability to surface them demonstrates its effectiveness at diagnosing issues.

Our expert interviews highlighted the benefits of the maturity model and the framework's potential to increase organizational transparency. E3 voiced that \textit{``it is useful to [...] [assure] AI system against goals that the organization set out for the system in the first place because it provides motivation for organizations to improve their maturity''}.
E1 expressed that if adopted industry-wide, the framework could \textit{``promote trust''} since there is currently \textit{``general obscurity of what organizations are doing''}.

Our framework's formal consideration of assurance in the ideation stage was identified as a novel and valuable contribution of our framework (E3).
Indeed, from our pilot applications, we found that it is crucial to conduct assurance in the early stages of the AI lifecycle for two reasons.
First, conducting early assurance enables detection of potential problems when they are typically easier to address.
In one pilot application, despite the end goal of supporting the day-to-day work of the end-users, they were not adequately consulted in value proposition, resulting in \textit{``building something that no one was asking for''} (S1).
As a result, this AI system was poorly received during testing and validation and never deployed.
Carrying out AI assurance in value proposition would have resulted in a less costly failure.

Second, conducting assurance early helps address lost and changing information.
In both pilot applications, there was stakeholder turnover: the team composition had changed over time, leading to knowledge gaps. 
For the business document tagging model, S4 was brought in later to address modeling difficulties and was not involved in the initial value proposition.
For the housing resource allocation model, S5 revealed that there were three different iterations of teams developing the AI system.
S9 said it \textit{``felt a little messy in [their] mind''} to answer questions about work in AI lifecycle stages that they had done a long time ago. 

Finally, E4 and E5 emphasized the importance of an ongoing partnership between assurance professionals and the organization developing the AI system to ensure the effectiveness of the process. The experts suggested clearly defining how assurance professionals will support the organization in progressing in maturity and securing the organization’s commitment to improving its practices when lower maturity levels are identified in specific areas.
Together, our findings show that our AI assurance framework is necessary and sound, and offer initial evidence that supports its usability and effectiveness.
At the same time, we identify key challenges for ensuring its continued applicability and impact, including the need for flexibility, appropriate timing of assurance, and the general constraints posed by incentives and limited resources.

\section{Discussion}

In this paper, we propose a third-party AI assurance framework.
We envision assurance as a mechanism to improve transparency, enable early risk detection, and provide generally applicable best practices and standards.
We focus in particular on \emph{third-party} assurance as it can reduce conflicts of interest, thereby increasing the credibility of the assurance mechanism.
We begin by identifying five \emph{design requirements} for a usable and effective assurance process, and highlight how these requirements have not been satisfied in the existing suite of RAI tools.
Then, we propose a \emph{prototype} of an AI assurance framework that satisfies our design requirements.
We conduct \emph{early testing} of our framework by applying it to two real-world AI systems and conducting expert interviews to verify the necessity, demonstrate the soundness, and provide initial evidence of the usability and effectiveness of our framework.
Our validations show positive first results that the framework enables structured third-party inquiry into an AI system, is accessible and learnable to professionals with domain expertise, and surfaces risks throughout the AI lifecycle.

By following the standard steps of design, and in particular, by conducting early testing, we are able to uncover the benefits and barriers to real-world adoption of our framework specifically, demonstrating the importance of conducting empirical assessments of RAI tools and following an iterative design process.
Our findings show that the framework is novel as an end-to-end artifact defining best standards and conducting third-party assurance.
While we identify some barriers to usability and effectiveness in our early validations, we believe that in spite of these, the framework itself is valuable as a first step toward filling an important gap in the literature.
Now, we reflect on two challenges that surfaced through our validations and conclude with limitations and future work.




\xhdr{Flexibility in the Assurance Process}
Our validations show that applying the AI assurance framework in practice requires substantial flexibility across use cases and continued adaptation as the AI landscape evolves. 
Across both pilot applications and expert interviews, participants emphasized that assurance questions need to be tailored to specific organizational contexts, stakeholder roles, and stages of the AI lifecycle.
This tailoring may entail rephrasing questions towards specific stakeholders (especially non-technical ones), determining which stakeholders to put more emphasis on at different stages of the AI lifecycle, and even tightening the definitions we provided for stakeholder groups and stages.

Importantly, this flexibility should be understood as a requisite skill for AI assurance professionals.
To apply the framework rigorously, assurance practitioners must exercise judgment in adapting its components while preserving its underlying structure and intent. 
Developing this capacity requires targeted training and professionalization~\cite{lam2024framework, kijewski2024rise, costanza2022audits, raji2022outsider},
and early efforts have begun to emerge~\cite{forhumanity2024welcome, algorithmicbiaslabndai}.
Our findings add \emph{specificity} to the skills required of assurance professionals (e.g. rephrasing questions for non-technical stakeholders) and thus to the competencies that should be covered in assurance training and certification.

Beyond individual practitioner expertise, our findings also highlight the need for plans to regularly update our framework, and RAI tools more broadly, as the AI development landscape changes. 
While our framework was intentionally designed as a \emph{blueprint} to accommodate such changes, our validations point to the importance of more explicit processes for monitoring and updating assurance frameworks over time, paralleling the recognition that AI systems themselves require continuous oversight~\cite{nist2023artificial, kuehnert2025who}.
We encourage researchers and practitioners to treat the maintenance of RAI tools as an ongoing responsibility rather than a one-time design task.

\xhdr{Creating Incentives to Undertake Assurance}
Our findings highlight that the incentives for undertaking such a process are not currently present.
Future work should explore the costs, organizational burdens, and scalability limitations associated with assurance as a means for understanding how assurance can be incentivized.

As a first step, we propose one mechanism to decrease the cost of assurance, which in turn, may encourage companies to undertake assurance:
{conducting the process in  the early stages of the AI lifecycle, especially in value proposition.
While assurance generally emphasizes evaluations that continue throughout the lifecycle of a system~\cite{ashmore2021assuring, lewis2023audit, robbins2024ai}, 
and while it would be ideal to conduct assurance at every stage of the AI lifecycle, our validations show that in a resource-constrained environment, assurance might be \emph{most} valuable when conducted early. Early assurance can detect potential issues proactively and result in a lower likelihood of lost information (e.g. through people leaving an organization).
Although the FAccT community has documented early-stage failures of AI systems~\cite{raji2022fallacy, godbole2025ai, faggella2018enterprise, coston2023validity} and proposed tools for the value proposition stage~\citep[e.g.][]{kawakami2024situate, yildirim2023investigating}, our findings underscore the need for \emph{external evaluations} during value proposition to prompt organizations to surface and mitigate problems before they become more severe and costly.
Consequently, we recommend that 
the community focus on developing evaluations targeting the value proposition stage of AI development.

Another key concern is that organizations may use the assurance framework and other evaluations as mechanisms for performative compliance~\citep[e.g.][]{kijewski2024rise, ferron2016does, power1997audit, bruen2010system}.
Towards this end, we also recommend that future work explore the extent to which mechanisms such as the periodic public disclosure of assurance reports can increase accountability by introducing reputational and market-based pressures on organizations.

\xhdr{Limitations and Future Work}  
First, in our pilot applications, we did not collect information on the outcomes of the AI system, such as the data used, model performance recorded, or documentation created.
This limitation reflects a broader challenge in AI assurance: outcomes are often difficult to access~\cite{mokander2021ethics}, especially for third-parties~\cite{costanza2022audits}, absent legal requirements~\cite{casper2024black}. 
Future evaluations that include greater access to system artifacts and outcomes would allow for deeper validation of the framework's ability to assess both outcomes and downstream effects. 
Second, our empirical validation relied on a small number of AI systems and expert interviews, and we recruited experts through convenience and snowball sampling.
We acknowledge the resulting limited generalizability of our findings. 
Future work can strengthen our validation by interviewing additional stakeholders in the AI systems we used for our pilot applications (such as impacted communities), conducting pilot applications on more AI systems, and obtaining more expert assessments.
Third, our validations are formative and exploratory rather than longitudinal. 
We propose that a key direction for future work is to develop rigorous and longitudinal evaluations of this AI assurance framework. 
Studies should involve applying the framework throughout the lifecycle of the AI system, implementing the recommendations that come from the assessments, and tracking changes in processes and outcomes overtime. 
Evaluations comparing teams that use the framework against those who rely on existing practices would further help to assess the framework's added value. 


\section{Endmatter Statements}

\subsection{Generative AI Usage Statement}

During manuscript preparation, the authors used generative AI tools (specifically, ChatGPT, OpenAI, GPT-4–class models) for limited editorial support, including grammar and style editing and sentence-level clarity improvements. 
Authors also used generative AI as one of many mechanisms for finding related work.
Generative AI tools were not used to generate original content, analyses, arguments, findings, or interpretations. 
All academic contributions including study design, data collection, analysis, interpretation, and writing were produced by the authors. 
The authors take full responsibility for the originality, accuracy, and integrity of the manuscript.

\subsection{Acknowledgments}
This research was funded by the Digital Transformation and Innovation Center at Carnegie Mellon University, sponsored by PricewaterhouseCoopers LLP (PwC) and the National Institute of Standards and Technology (\url{ror.org/05xpvk416}) under Federal Award ID Number 60NANB24D231 and Carnegie Mellon University (\url{https://ror.org/05x2bcf33}) AI Measurement Science and Engineering Center (AIMSEC). 
RMK is additionally supported by NSERC PGS D fellowship (599255).

\subsection{Ethical Considerations Statement}\label{sec:ethics}
To preserve anonymity of interview participants, we assured interviewees that their participation was voluntary, they could decline to answer interviewer questions, and their responses would be kept anonymous. 
For sensitive or potentially identifying interview quotes, we exclude participant IDs to preserve anonymity. 
To mitigate the risk that participating organizations are identified, we limit the amount of detail we provide about each organizations' practices.

\bibliographystyle{ACM-Reference-Format}
\bibliography{references}


\clearpage
\appendix
\section{Stakeholder and Stage Descriptions}

Table~\ref{tab:stakeholders} contains the definitions of each of the stakeholders we consider in our AI assurance framework.

\begin{table*}[]
\centering
\scriptsize
\begin{tabular}
{cp{0.8\linewidth}}
\toprule
\textbf{Role}
 & \textbf{Definition} \\ \midrule
\textbf{Leaders}           &  This role makes strategic decisions about the goals, objectives, and missions of the organization and the high-level policies to implement those goals. Typically, leaders are at the top of the organizational hierarchy or structure and include roles such as C-suite executives, and leaders of units or teams who are delegated specific responsibilities. In addition to C-suite executives, these roles can include senior management, unit or group leaders, policy leaders, or team leaders.  \\ \midrule
\textbf{Designers}         &  This role oversees the reframing of business and policy goals into a technical problem, aligning these problems with stakeholder needs in a particular context of use, and envisioning the user experience. Example roles include domain experts, AI designers, product managers, compliance experts, human factors experts, UX designers, and others who are familiar with doing cross-functional work.  \\ \midrule
\textbf{Developers}        &  This role uses programming or other technical knowledge to prepare, curate, and engineer data and to train AI models to perform the specified task. Example roles include data engineers, modelers, model engineers, data scientists, AI developers, software engineers, and systems engineers.  \\ \midrule
\textbf{Deployers}         &  The deployer role verifies and validates the model beyond the training and test data, pilots the model, checks compatibility with existing systems, and collaborates with designers to evaluate the user experience. Roles include testing and evaluation experts, auditors, impact assessors, and system integrators.   \\ \midrule
\textbf{End-users}         &  This role represents any individual who utilize the output of the AI model to contribute to an organizational goal (for example, making decisions or automating workflows and practices).  \\ \midrule
\textbf{Impacted Communities} & This stakeholder group represents any individual or community impacted directly or indirectly by the AI system’s operation and their advocates and representatives. Roles include groups that may be harmed by the model, advocacy groups, and the broader public.    \\ \bottomrule
\end{tabular}
\caption{Role definitions for the major stakeholders in an AI system. The wording is largely borrowed from the work of~\citet{kuehnert2025who}.}
\label{tab:stakeholders}
\end{table*}

\xhdr{Stage Descriptions}
Table~\ref{tab:stages} contains the definitions of each of the stages we consider in our AI assurance framework.

\begin{table*}[]
\centering
\scriptsize
\begin{tabular}
{cp{0.8\linewidth}}
\toprule
\textbf{Stage}
 & \textbf{Definition} \\ \midrule
\textbf{Value Proposition}           &  Value Proposition refers to a series of early investigations into the problem the AI system is designed to solve, what the business and policy requirements are, and whether including an AI component within the broader decision-making system leads to net-benefit over the status quo. \\ \midrule
\textbf{Problem Formulation}         &  Problem Formulation refers to the translation of business needs and requirements to technical choices, what types of methodologies will be used, and how the overall system will incorporate the AI system.   \\ \midrule
\textbf{Data Collection}        &  Data collection involves selecting, collecting, or compiling data to train and/or validate an AI model, and it involves making choices---or implicitly accepting previously made choices---about how to sample, label, and link data.  \\ \midrule
\textbf{Data Processing}         &  Data processing refers to the steps taken to make data usable by the ML model---for example, cleaning, standardizing, or normalizing data, including feature engineering. \\ \midrule
\textbf{Statistical Modeling}    &  Statistical Modeling consists of deciding what type of statistical model(s) to fit to the data, and how to perform the fitting. The latter requires choosing the learning rule and loss function, regularizers, hyperparameters' values, model selection methodology, and model selection metrics to be used. \\ \midrule
\textbf{Testing and Validation} & Testing and Validation refers to the process by which a model is determined to be performing well enough for deployment. Some major choices include metrics to evaluate the model on, and subpopulations for which the model should be evaluated.   \\ \midrule
\textbf{Deployment}         &  Deployment refers to the process of deploying the model into a larger decision system.  \\ \midrule
\textbf{Monitoring}         &  Monitoring refers to how a model’s behavior is recorded and responded to over time to ensure there is no significant or unexpected degradation in performance and use over time.  \\ \bottomrule
\end{tabular}
\caption{Definitions for major stages in the AI lifecycle. The wording is largely borrowed from the work of~\citet{kuehnert2025who}.}
\label{tab:stages}
\end{table*}


\section{AI Assurance Framework Components}

\subsection{Responsibility Assignment Matrix}

Table~\ref{tab:raci} shows a high-level schematic of what a responsibility assignment matrix looks like.
Each cell of the matrix contains a level of responsibility, or involvement, that the stakeholder had at that stage of the AI lifecycle. 
We do not fill out the levels of responsibility as they may differ depending on the AI system; however, we believe that there are some (stakeholder, stage)-intersections that have higher levels of responsibility across many use cases (such as (Developers, Statistical Modeling)).
These intersections informed where we added more questions to the interview protocol.
However, since we conducted semi-structured interviews, the participants' responses also informed the stages of the AI lifecycle in which we probed the participants for additional information.

\begin{table*}[]
    \tiny
    \centering
    \begin{tabular}{c|c|c|c|c|c|}
        \toprule
         & Value Proposition and Problem Formulation & Data Collection and Processing & Statistical Modeling & Testing and Validation & Deployment and Monitoring \\
         \midrule
         Leaders &&&&& \\
         \midrule
         Designers &&&&& \\
         \midrule
         Developers &&&&& \\
         \midrule
         Deployers &&&&& \\
         \midrule
         End-users &&&&& \\
         \midrule
         Impacted Communities &&&&& \\
         \bottomrule
    \end{tabular}
    \caption{High-level schematic of a responsibility assignment matrix. The leftmost column contains the key stakeholder groups of an AI system. The top row contains the stages of the AI lifecycle. Each (stakeholder, stage)-intersection in the matrix would contain a level of responsibility, or involvement, that the stakeholder had at that stage of the AI lifecycle.}
    \label{tab:raci}
\end{table*}

\subsection{Interview Protocol}\label{sec:app-stakeholder-interview-protocol}

We began with an introduction to the project and confirmed participants' consent in participating in this project.

\textbf{Position and Experience}
\begin{itemize}
    \item Can you tell me more about you, your role at your company, and your academic and professional background? 
    \item Which AI stakeholder group most closely captures your role in this project?
    \item What is your role or main focus? (e.g., leaders, designers, developers, deployers, end-users, and impacted communities, etc.)
    \item Can you tell me a little bit about your role in the project? 
    \item How many years of experience do you have working with AI systems as a leader / developer / deployer / etc.?
    \item What was the purpose of the AI system?
    \item What is your impression of how AI is currently governed in your organization? 
    \begin{itemize}
        \item By this, I mean how is it decided what to build?
        \item How to build an AI system?
        \item How to evaluate an AI system?
        \item Who has oversight?
        \item Who takes accountability when something goes wrong?
    \end{itemize}
    \item In what ways can your organization improve its governance of the technology?
    \item Did you have the chance to read the pre-reading document (containing the definitions for the stakeholders and stages)?
    \item Do you have any questions?
\end{itemize}

\textbf{General Questions}

Note: these questions are asked to every participant in every stage of the AI lifecycle.

Note: we customized the protocol based on the people we were interviewing, and the project we were talking about. For example, if we were interviewing a less technical stakeholder, we would further define terms such as ``Value Proposition'' within the context of the specific project.

\begin{itemize}
    \item What do you think are the key questions someone in your role as a [stakeholder group] should consider at [stage of AI lifecycle]?
    \item How would you describe the responsibility assigned to your role as a [stakeholder group] in [stage of AI lifecycle] for [project name]?
    \item What specific tools or processes did you use to discharge your responsibilities?
    \item Are there any other tools or processes that you know of that you could use to discharge your responsibilities?
    \item How do you think responsibilities should be allocated to different stakeholder groups at  [stage of AI lifecycle]?
    \item How would you change what was done in [stage of AI lifecycle] for [project name]?
    \item Who do you think should do the work in [stage of AI lifecycle]?
    \item Who do you think should take ownership of any of the key decisions made in  [stage of AI lifecycle]
    \item Who do you think should be involved in  [stage of AI lifecycle]?
\end{itemize}

\textbf{(Stakeholder, Stage)-Specific Questions}

A ``Main Question'' is asked if they are not already touched upon by the participant in the answers to the above questions. A ``Follow-up and Optional Question'' is asked if time permits.

\textbf{\textit{Value Proposition and Problem Formulation}}

\textit{Leaders}

\begin{itemize}
    \item Main Question: Did you formally assess the value of this AI system?
    \item Follow-up and Optional Questions
    \begin{itemize}
        \item If yes, can you describe that process?
    \end{itemize}
    \item Main Question: In deciding to move forward with this AI system, can you describe some of the benefits and risks you considered (or will consider)?
    \item Follow-up and Optional Questions
    \begin{itemize}
        \item Did you consider the technical feasibility and resources required for the AI system?
        \item If yes, what aspects did you consider?
        \item What are some non-AI alternatives that you considered?
        \item Which stakeholders did you talk to during value proposition and problem formulation?
        \item How did you take their opinion into account?
        \item Did you consider how the AI system would fit into existing organizational contexts and existing practices?
        \item If yes, what aspects did you consider?
    \end{itemize}
    \item Main Question: Can you describe how you have documented all the processes in value proposition and problem formulation?
    \item Follow-up and Optional Questions
    \begin{itemize}
        \item What was the purpose?
        \item What impact did you expect?
        \item How did (will) you assess this impact?
        \item How did (will)  you consider risks and benefits?
        \item Why did you decide to begin to work on an AI system for homelessness?
        \item Did you decide that the AI system provides a net benefit?
        \item If yes, how did you come to this conclusion?
    \end{itemize}
\end{itemize}

\textit{Designers}
\begin{itemize}
    \item Main Question: Please describe how the real-world problem that the AI system addresses was explored, understood, and documented.
    \item Main Question: How did you translate the real-world problem into an AI problem?
\end{itemize}

\textbf{\textit{Data Collection and Processing}}

\textit{Leaders}
\begin{itemize}
    \item Main Question: Can you describe how you determined what your data needs and requirements were / are for the AI system?
    \item Main Question: Can you describe how you assess whether the data use is complying with legal, regulatory, and ethical requirements?
    \item Main Question: Can you describe your plan to respond to any data breaches or unauthorized use of the data?
    \item Main Question: Can you describe how you decided who the data will be made accessible to others and to what extent?
    \item Main Question: How do you (plan to) assess the suitability of the data sources used to create this AI system?
    \item Follow-up and Optional Questions
    \begin{itemize}
        \item What data sources do you use for this model?
        \item How did you decide which data sources to use for this project?
        \item Was any data collected specifically collected for the purpose of this AI system?
        \item Can you describe if there are any data quality checks in place?
        \item Why are the tools you use the ones you use?
        \item What processes are in place to store, update, and use datasets?
    \end{itemize}
\end{itemize}

\textit{Designers}
\begin{itemize}
    \item Main Question: Can you describe how you determine whether the data you have is suitable for the problem at hand?
    \item Main Question: Describe how the data was stored, maintained, and updated and why you made those decisions.
    \item Who was storing, maintaining, and updating the data?
    \item Main Question: Which stakeholders did you talk to during data collection and processing?
    \item Follow-up and Optional Questions
    \begin{itemize}
        \item How did you take their opinion into account?
    \end{itemize}
    \item Main Question: How have you documented all the processes in data collection and processing?
    \item Follow-up and optional questions
    \begin{itemize}
        \item How do you determine if and when you will need to acquire additional data sources?
        \item If yes, what is this process?
        \item What are the tradeoffs that are explored in this process?
        \item Can the model lead to data drift?
        \item If yes, how do you plan on addressing this?
    \end{itemize}
\end{itemize}

\textit{Developers}
\begin{itemize}
    \item Main Question: Can you describe how you determined what your data needs and requirements were / are for the AI system?
    \item Main Question: Are there specific measures or metrics you use to assess data quality? And can you describe how you chose them?
    \item Follow-up and Optional Questions
    \begin{itemize}
        \item Do you look at the representativeness of different groups?
        \item If yes, what groups do you look at?
        \item Why do you believe that there is a causal relationship between your data and the target of prediction?
    \end{itemize}
    \item Main Question: What data cleaning and engineering processes did you use?
    \item Follow-up and Optional Questions
    \begin{itemize}
        \item Why did you choose to use these processes?
        \item How did you handle missing values?
        \item How did you handle outliers?
        \item How did you handle redundant instances?
    \end{itemize}
\end{itemize}

\textit{\textbf{Statistical Modeling}}

\textit{Developers}
\begin{itemize}
    \item Main Question: Describe how you processed the data and why you made those decisions.
    \item Main Question: Can you describe the process you used to select the metrics to evaluate your model on?
    \item Main Question: Describe how the development team chose the model and related training procedures?
    \item Main Question: Which stakeholders did you talk to during statistical modeling?
    \item Follow-up and Optional Questions
    \begin{itemize}
        \item How did you take their opinion into account?
    \end{itemize}
    \item Main Question: How have you documented all the processes in statistical modeling?
    \item Follow-up and Optional Questions
    \begin{itemize}
        \item Why did you choose the model you chose?
        \item Why not a different model (maybe one that is more interpretable)?
        \item How did you train your model?
        \item What is the objective function of the model?
        \item How much improvement over existing benchmarks does the AI system perform?
        \item How does the model perform across different groups?
        \item How do you measure performance?
        \item Why do you choose to measure performance in this way/
        \item How generalizable is the model?
        \item How did you measure generalizability?
        \item Why did you choose to measure generalizability in this way?
    \end{itemize}
\end{itemize}

\textit{\textbf{Testing and Validation}}

\textit{Developers}
\begin{itemize}
    \item Main Question: Can you describe what metrics and measures  you use to evaluate the trained model in real-world contexts?
    \item Follow-up and Optional Questions 
    \begin{itemize}
        \item How did you decide to use those metrics?
    \end{itemize}
\end{itemize}

\textit{Deployers}
\begin{itemize}
    \item Main Question: Can you describe how you determine who benefits and who is harmed by the AI system in real-world usage?
    \item Main Question: Which stakeholders did you talk to during testing and validation?
    \item Follow-up and Optional Questions
    \begin{itemize}
        \item How did you choose those stakeholders?
        \item How did you take their opinion into account?
    \end{itemize}
    \item Main Question: How have you documented all the processes in testing and validation?
\end{itemize}

\textit{\textbf{Deployment and Monitoring}}

\textit{Leaders}
\begin{itemize}
    \item Main Question: How much improvement over existing baselines does the AI system perform?
    \item Follow-up and Optional Questions
    \begin{itemize}
        \item What are the existing baselines?
        \item How do you measure improvement?
        \item Why did you choose to measure improvement this way?
    \end{itemize}
    \item Main Question: Can you describe how you ensure that someone is accountable for monitoring the AI system?
    \item Follow-up and Optional Questions
    \begin{itemize}
        \item Who is monitoring the AI system?
        \item What would cause you to shut the AI system down?
        \item How do you track and manage incidents involving this AI system?
        \item How are impact assessments performed?
        \item How often are they performed?
        \item What are the risk thresholds that are in place?
    \end{itemize}
\end{itemize}

\textit{Designers}

\begin{itemize}
    \item Main Question: Can you describe how you design and document the integration of the new AI system into larger workflows and systems?
    \item Main Question: Can you describe the guidance you provided to the end-user about how to use the AI system?
    \item Follow-up and Optional Questions
    \begin{itemize}
        \item Is the documentation stored anywhere?
        \item How is the output of your model used?
        \item Why do you choose to use the output in this way?
        \item How do you measure the success of your model / the model-human collaboration?
        \item What are the criteria that define success?
        \item Why do you choose to measure success in the way that you measure it?
    \end{itemize}
\end{itemize}

\textit{Deployers}

\begin{itemize}
    \item Main Question: Based on the evaluations you did, can you describe how you determined that this model can be moved to the stage of deployment?
    \item Follow-up and Optional Questions
    \begin{itemize}
        \item Were you in the rollout plan (e.g. staged rollout)?
        \item How did you train the end-users to use the model?
    \end{itemize}
    \item Main Question: What was your process for ensuring compliance with relevant regulatory requirements?
    \item Main Question: Can you describe how you are monitoring the performance of the model?
    \item Follow-up and optional questions
    \begin{itemize}
        \item What metrics are you using?
        \item Why are you using these metrics?
        \item Can you describe how you chose these metrics?
        \item How do you know when to update the model?
    \end{itemize}
    \item Main Question: Can you describe the primary, secondary, and unintended uses of the model? How did you consider them during model deployment?
    \item Main Question: How do you know when to roll back the model?
    \item Main Question: How do you decide how much of the methodology, datasets, code, and impact measurements can be made publicly available?
    \item Main Question: Which stakeholders did you talk to during deployment and monitoring?
    \item Follow-up and Optional Questions
    \begin{itemize}
        \item How did you take their opinion into account?
    \end{itemize}
    \item Main Question: How have you documented all the processes in deployment and monitoring?
\end{itemize}

\textit{End-Users}
\begin{itemize}
    \item Main Question: How do you use the output of the model to make a decision?
    \item Follow-up and Optional Questions
    \begin{itemize}
        \item Can you describe how the model integrates into your daily workflow?
        \item Were you involved in the design, field-testing and trial-runs surrounding the model? If so, how?
    \end{itemize}
\end{itemize}

\textbf{\textit{Impacted Communities}}

Note: Impacted community members may have less familiarity with the model; thus, we ask the following questions that are not specific to any stage of the AI lifecycle.

\textit{General Questions}
\begin{itemize}
    \item Main Question: How do you think this model can impact you?
    \item Main Question: What role do you play in assessing the impact of the model in your community?
    \item Follow-up and Optional Questions
    \begin{itemize}
        \item How can you provide input about the model to [your organization]?
    \end{itemize}
\end{itemize}

\textbf{Summary and Final Thoughts}

\begin{itemize}
    \item What do you think about the Responsibility Assignment Matrix framework?
    \item Was there anything that we did not discuss that you would like to share or address?
\end{itemize}


\subsection{Maturity Matrix}\label{sec:app-maturity-matrix}


Each ``category'' of the maturity matrix contains a description or questions that further explain what the category is meant to capture.
The ``additional notes'', highlighted in blue, clarify the content within certain cells.

\includepdf[pages=-, landscape=true]{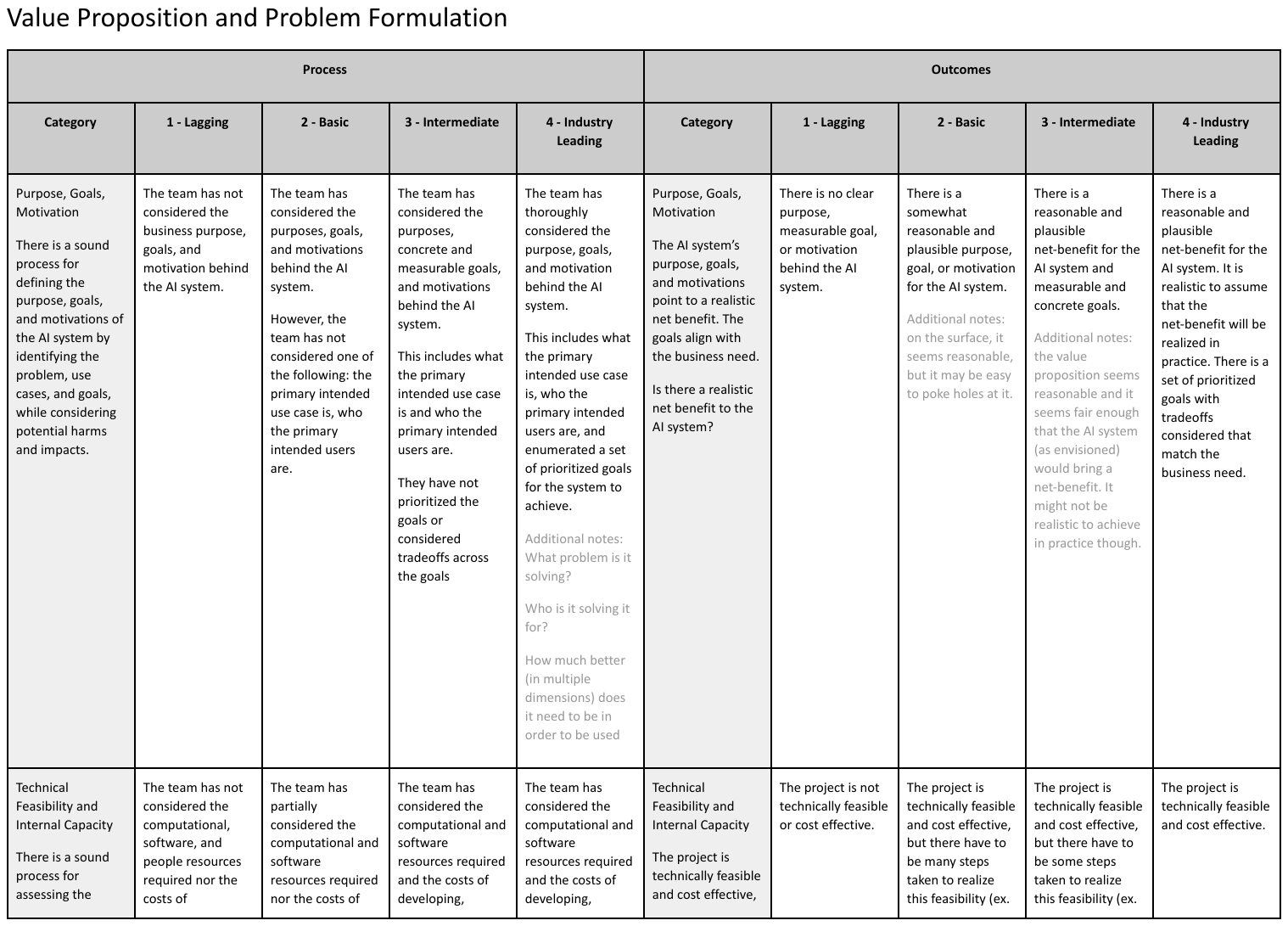}

\subsection{Assurance Report Format}\label{sec:app-report-format}

The final assurance report is produced by consolidating the findings of prior phases and summarizing them into assurance ratings described earlier. We provide one of the four assurance ratings for the AI system overall, and for each stage of the AI lifecycle.
\begin{enumerate}
    \item Inadequate access to perform the evaluation.
    \item Adequate access, lack of compliance with baseline industry expectations.
    \item Adequate access; adequate compliance with baseline expert expectations but lack of evidence of compliance with state-of-the-art and best practices.
    \item Adequate access, adequate compliance with best practices, and state-of-the-art industry standards.
\end{enumerate}

To arrive at the overall assurance rating, we use the following criteria.
\begin{itemize}
    \item If any stage has an assurance rating of 1, then we give an overall assurance rating of 1.
    \item If any stage has an assurance rating of 2, then we give an overall assurance rating of 2.
    \item If any stage has an assurance rating of 3, and all the other stages have an assurance rating of 3 or higher, then we give an overall assurance rating of 3.
    \item If every stage has an assurance rating of 4, then we give an overall assurance rating of 4.
\end{itemize}

To arrive at the assurance rating for any individual stage, we use the following rules.
\begin{itemize}
    \item If any subcategory (for both process and output) cannot be rated due to inadequate access, then we give an assurance rating of 1 for the stage.
    \item If any subcategory rating (for both process and output) is 1, then we give an assurance rating of 2 for the stage.
    \item If any subcategory rating (for both process and output) is 2, and all the other subcategories have a subcategory rating of 2 or higher, then we give an assurance rating of 3 for the stage.
    \item If every subcategory rating (for both process and output) is 3 or higher, then we give an overall assurance rating of 4 for the stage.
\end{itemize}

For more details on the structure of the report, see below.

\includepdf[pages=-]{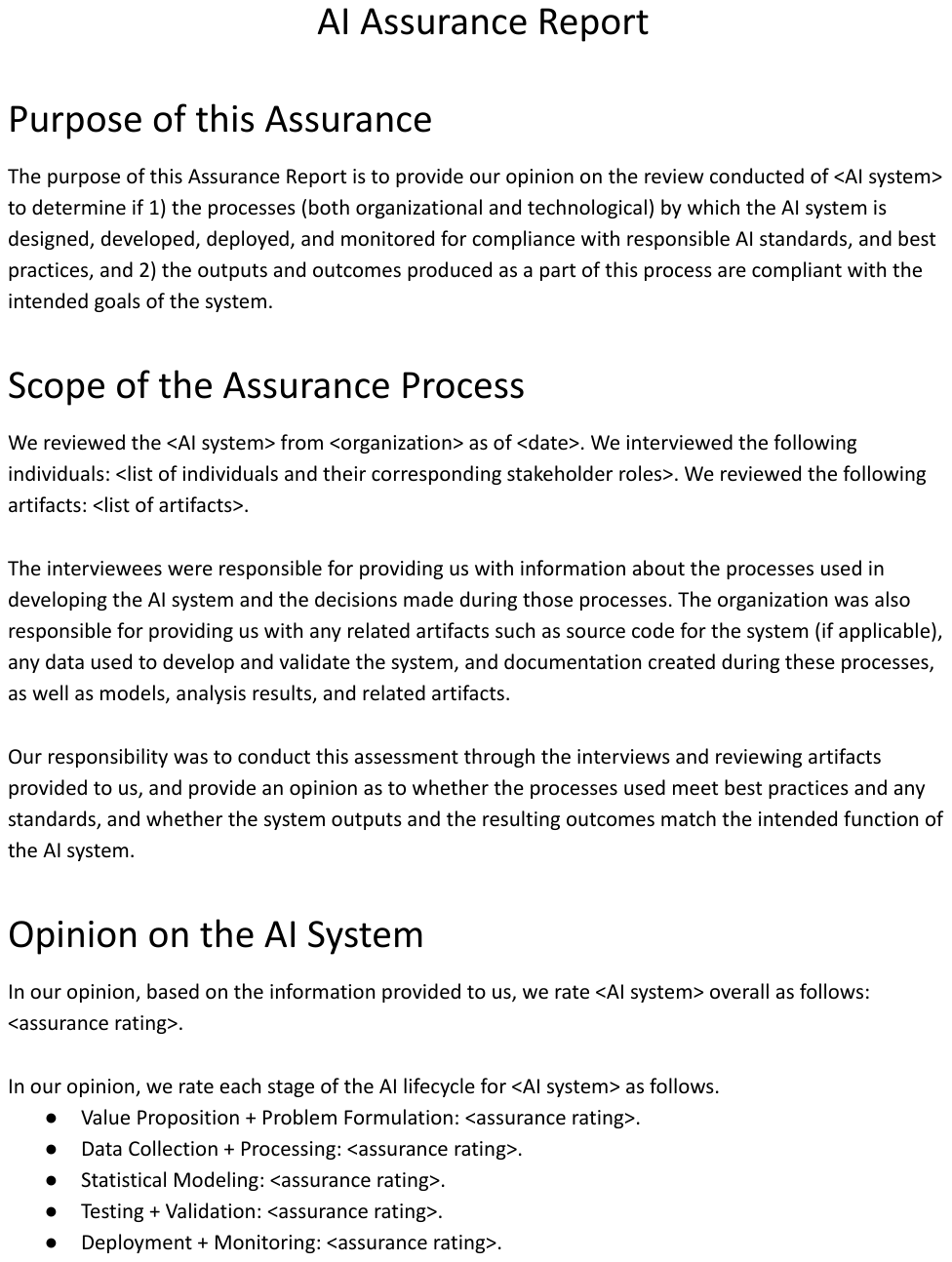}

\section{AI Assurance Framework Validation}

\subsection{Expert Validation Interview Protocol}\label{sec:app-expert-interview-protocol}

We began with an introduction to the project and confirmed participants' consent in participating in this project.

\textbf{Position and Experience}
\begin{itemize}
    \item Can you tell me more about you, your role at your company, and your academic and professional background as it relates to AI assurance, audits, or evaluations?
    \begin{itemize}
        \item Please describe your level of familiarity with AI assurance, audits, and evaluations.
    \end{itemize}
\end{itemize}

\textbf{Verifying Needs}

In this section, we focus on quickly establishing whether experts perceive a substantive need for an assurance framework. Then we explore how the participant perceives current assessment practices.
\begin{itemize}
    \item How are AI systems currently assessed in your organization or field?
    \begin{itemize}
        \item Are there any standards or shared practices that you or your colleagues rely on?
    \end{itemize}
\end{itemize}

\textbf{Presentation}

We give a 10-15 minute presentation of our AI assurance framework.
\begin{itemize}
    \item Are there any other third-party end-to-end frameworks, other than the one that has been presented, that help people conduct AI assurance?
    \begin{itemize}
        \item If yes: ask the questions based on comparisons to these frameworks (under subheader ``Questions About Framework in Comparison to Others'')
        \item If If no: ask the questions about the framework itself (under subheader ``Questions About Framework'')
    \end{itemize}
\end{itemize}

\textbf{Verifying Soundness}

In this section, we aim to quickly verify that the framework is coherent and contains no major gaps.

\begin{itemize}
    \item What is your assessment of the general approach?
    \begin{itemize}
        \item Are all major stages of the AI lifecycle included?
        \item Are all major stakeholders in an AI system included?
    \end{itemize}
    \item Are there any big conceptual gaps or critical steps that seem to be missing from this framework?
    \item Does anything feel redundant, confused, or misaligned with how AI systems are actually assessed?
    \item Do you have any other comments on the coherence and completeness of our framework?
\end{itemize}

\textbf{Verifying Usability}

In this section, we want to explore the usability of the framework.

\textbf{\textit{Questions About Framework}}
\begin{itemize}
    \item What is your assessment of the applicability of the framework?
    \begin{itemize}
        \item For what kinds of AI systems do you think this framework is applicable?
        \item For what kinds of AI systems do you think this framework is inapplicable?
        \item How applicable do you believe the framework is across different stages of the AI lifecycle?
    \end{itemize}
    \item What is your opinion on the ability of people who typically conduct assurance or AI audits to apply this framework?
    \begin{itemize}
        \item Do you think the framework requires deep technical expertise, contextual expertise, or both?
        \item Does this level of expertise feel too demanding, too light, or about right?
    \end{itemize}
    \item For people who typically conduct assurance or AI audits, how easy or hard do you believe it would be to quickly learn and use the AI assurance framework after first encountering it?
    \item What do you believe is the chance that someone could make an error in following the assurance process?
    \begin{itemize}
        \item What kinds of errors would they make?
        \item How severe would the errors be? 
        \item How easily can the user recover from the errors?
    \end{itemize}
\end{itemize}

\textbf{\textit{Questions About Framework in Comparison to Others}}

Please compare this framework vs. <other AI assessment or auditing process>, in terms of...

\begin{itemize}
    \item How does the applicability of this framework vs. <other AI assessment or auditing process> feel across different AI systems or contexts?
    \begin{itemize}
        \item For what kinds of AI systems do you think this framework is more applicable than <other AI assessment or auditing process>?
        \item For what kinds of AI systems do you think this framework is less applicable than <other AI assessment or auditing process>?
        \item Can you compare the applicability of the framework vs. <other AI assessment or auditing process> across different stages of the AI lifecycle?
    \end{itemize}
    \item How would you assess the ability of people who typically conduct assurance or AI audits to apply this framework vs. <other AI assessment or auditing process>?
    \begin{itemize}
        \item Do you think the framework requires deep technical expertise, contextual expertise, or both? Compared to <other AI assessment or auditing process>?
        \item Does this level of expertise feel too demanding, too light, or about right? Compared to <other AI assessment or auditing process>?
    \end{itemize}
    \item For people who typically conduct assurance or AI audits, how easy or hard do you believe it would be to quickly learn and use this AI assurance framework vs. <other AI assessment or auditing process> after first encountering it?
    \item What do you believe is the chance that someone could make an error in following this assurance process vs. <other AI assessment or auditing process>?
    \begin{itemize}
        \item What kinds of errors would they make? Compared to <other AI assessment or auditing process>?
        \item How severe would the errors be? Compared to <other AI assessment or auditing process>?
        \item How easily can the user of the AI assurance framework recover from the errors? Compared to <other AI assessment or auditing process>?
    \end{itemize}
\end{itemize}

\textbf{Verifying Effectiveness}

In this section, we want to have the participant assess whether the AI assurance framework works, and how it could lead to better outcomes. 

\textbf{\textit{Questions About Framework}}
\begin{itemize}
    \item What is your assessment of the potential of this framework to improve organizational practices surrounding designing, developing, and deploying an AI system?
    \begin{itemize}
        \item Suppose an organization were to follow the recommendations outlined in the framework. What do you see as the potential for the recommendations to improve organizational practices?
    \end{itemize}
    \item What is your assessment of the potential of this framework to improve AI system outcomes?
    \begin{itemize}
        \item Can you describe the extent to which following the recommendations outlined in the framework would improve organizational practices?
        \item Ex. Would using this framework help shut down AI systems that should not be built in the first place earlier?
    \end{itemize}
    \item What do you think the overall impact will be if this AI assurance framework becomes industry standard?
\end{itemize}

\textbf{\textit{Questions About Framework in Comparison to Others}}
\begin{itemize}
    \item Please compare applying this framework vs. other AI assessment or auditing frameworks you are familiar, in terms of their potential to improve:
    \begin{itemize}
        \item Organizational practices surrounding designing, developing, and deploying an AI system
        \item AI system outcomes
    \end{itemize}
    \item What do you think the overall impact will be if this AI assurance framework becomes industry standard?
\end{itemize}

\textbf{Deep Dive into the Framework}

In this section, we want to take a closer look at a few of the stages and categories to assess how well we covered the stages and processes. 

Note: we only ask questions from this section if there is time.

\textbf{\textit{Subcategories in the Maturity Matrix}}

We take a sample AI lifecycle stage (Testing and Validation).
\begin{itemize}
    \item Do the subcategories within each AI lifecycle stage accurately capture the major steps that matter?
    \begin{itemize}
        \item Anything unnecessary, missing, or misclassified?
    \end{itemize}
\end{itemize}

\textbf{\textit{Assertions About Lagging / Basic / Intermediate / Industry-Leading}}

We take a sample row / rows from the maturity matrix (Organizational Feasibility and Models Selection Methodology---Requirement Satisfaction).
\begin{itemize}
    \item Does the separation of process and outcome make sense?
    \item Do the descriptions of lagging, basic, intermediate, and leading maturity seem accurate?
    \item Does anything strike you as unrealistic, too strict, or too lenient?
\end{itemize}

\textbf{Summary and Final Thoughts}
\begin{itemize}
    \item Was there anything that we did not discuss that you would like to share or address? 
\end{itemize}





\end{document}